\documentclass[12pt,twoside,a4paper]{article}
\author{Jan-Markus Schwindt$^1$, Christof Wetterich$^2$}
\date{}
\title{Dark energy cosmologies for codimension-two branes}

\begin{document} 
\maketitle 

\centerline{\small\it $^1$Institut f\"ur Physik, Universit\"at Mainz, Staudingerweg 7,
55128 Mainz}
\centerline{\small\it \quad E-mail: Schwindt@thep.physik.uni-mainz.de}
\vspace{0.3cm}
\centerline{\small\it $^2$Institut f\"ur Theoretische Physik, Universit\"at
Heidelberg,
Philosophenweg 16, 69120 Heidelberg}
\centerline{\small\it \quad E-mail: C.Wetterich@thphys.uni-heidelberg.de}
\vspace{0.7cm} 

\begin{abstract}
A six-dimensional universe with two branes
in the ``football-shaped" geometry leads to an almost realistic cosmology.
We describe a family of exact solutions with time dependent characteristic size of internal
space. After a short inflationary period the late cosmology is either of quintessence type
or turns to a radiation dominated Friedmann universe where the cosmological constant appears
as a free integration constant of the solution. The
radiation dominated universe with relativistic 
fermions is analyzed in detail, including its dimensional reduction.  
\end{abstract}

\section{Introduction}
The cosmological constant problem is the question why the observed curvature of spacetime
is so much smaller (by 120 orders of magnitude) than the naively guessed gravitational scale.
In higher dimensional theories, such as Kaluza-Klein theories or braneworlds, the 
total curvature $R^{(4+d)}$ can be divided into three parts: the four-dimensional (4D) curvature 
$R^{(4)}$ of the visible universe, the curvature $R^{(d)}$
of internal space (the bulk) and a term for the warping. The total curvature is expected
to be of the order of the fundamental scale of the theory. In this context, the 
cosmological constant problem can be formulated in a different way: Why is only such a
small part of $R^{(4+d)}$ contained in $R^{(4)}$?

In this context Rubakov and Shaposhnikov noted \cite{rusha} 
that in a six-dimensional pure
gravity theory $R^{(4)}$ is a free integration constant of the general solution 
and may have an arbitrary value,
no matter how large $R^{(6)}$ is. The same is true \cite{cw1} in 6D Einstein Maxwell
theory where two dimensions are compactified by magnetic flux \cite{rdss1}. The latter model
does not share the singularities appearing in the Rubakov-Shaposhnikov model and has
better stability properties \cite{cw1,rdss1,latin}. Its geometry corresponds to the bulk
of braneworlds (with codimension 2) which are extensively discussed today. 
The remaining question is then: why would a solution with very 
small $R^{(4)}$ be selected? 

A somewhat similar problem appears in standard four-dimensional cosmology. 
The total 4D curvature of the
universe consists of two terms: the 3D curvature of the spatial hypersurfaces 
and the Hubble parameter $H=\frac{\dot{a}^2}{a^2}$ which measures the time dependence
of these hypersurfaces and corresponds to the warping. 
Measurements of the CMB anisotropies show that the 3D curvature
is almost zero, at least much smaller than the Hubble parameter. Standard cosmology
implies then that this mismatch in scale between 3D curvature and 4D curvature was even 
much more pronounced in the past. This raises the 
question: Why does the 3D curvature carry only such a small part of the total 4D 
curvature? In this case a dynamical solution to the problem is known: inflation. During
inflation, the 3D curvature is exponentially driven to zero, while the 4D curvature
is essentially given by the energy density and pressure of the inflaton, which remains
almost constant.

One may ask the question if the cosmological constant problem as stated above, 
i.e. the question why the 4D curvature carries only such a small part of the total
higher dimensional curvature, may also have a dynamical solution. 
This could happen in two ways: The first possibility concerns solutions which approach
asymptotically for large time a static and
stable internal space and where the four-dimensional cosmological constant is a free integration
constant of the general solution. 
A very small value of this constant may be favoured by the dynamics of the
very early universe when the size and shape of internal space were strongly time-dependent and
the stable solution with fixed small $\Lambda _4$ could have been asymptotically approached.
Such a scenario would be the analogue of inflation as discussed above. 
A second possibility is that the internal space remains time-dependent even at late times. This 
may lead to quintessence-type \cite{rp} 
effective four-dimensional cosmologies where the ``dark energy"
density becomes small and time-dependent by virtue of a cosmic attractor solution, 
independently of the parameters of the model and the integration constants of the solution. 
We will discuss both possibilities
here, but restrict ourselves to a particular six-dimensional example.

The cosmological constant problem in six dimensions has recently been discussed
within the context of braneworlds 
\cite{fb,codim2,susy6d,nilles,ns}.
In this paper we present new cosmological solutions for the so-called ``football
shaped" model \cite{fb}. The two-dimensional internal space has axial symmetry
and two equal conical singularities at the two poles, giving it the shape of an American
football. It is stabilized by the magnetic flux of the six-dimensional gauge field.

It is known that this model exhibits static solutions where the 4D cosmological constant
appears as a free integration constant \cite{cw1}. In this paper we demonstrate that
there exists at least one cosmological solution which approaches a given static
solution asymptotically. This holds for arbitrary values of the integration constants
of the general static solution.
Beyond earlier work we investigate the full time dependence of the volume of internal
space and the 4D scale factor. We establish exact cosmological 
solutions of the 6D field equations.

The time-dependent size of internal space appears in four dimensions as 
a time-dependent scalar field which has substantial influence on cosmology. The 
shape of the corresponding scalar potential allows for a short period of inflation,
a stable ground state with or without cosmological constant, or alternatively for a 
cosmology with exponential potential quintessence. As a second difference to other work we
do not postulate that the usual matter is so much constrained on the brane that it
becomes ``invisible" from the bulk. Instead we investigate ``holographic branes" 
\cite{holo} where the 
wave functions of all particles have a normalizable tail in the bulk, or even have
a maximum there.
This requires the size of the bulk to be small. In consequence, the ``superheavy Kaluza-Klein
modes" decay early in cosmology and play no role for the radiation or matter 
dominated late universe.

This point of view surrounds all the recently discussed problems with
codimension-two branes and the cosmology appears to be quite realistic. 
Indeed, we find that the energy momentum tensor of the massless
fermions or gauge fields generates a usual Friedmann-Robertson-Walker
cosmology in the observed four-dimensional world. Together these features locate our
solutions much closer to reality than the other approaches recently discussed.

The paper is organized as follows: In section 2 we present cosmological solutions
of 6D Einstein-Maxwell theory with ``football shaped" internal space of time-dependent
size. From a four-dimensional perspective, 
this size appears as a scalar field $\phi$ whose potential
is determined by the bulk physics. It has in general a minimum and a maximum, and
we show that inflation as well as Friedmann cosmology can be obtained in this context.
In section 3 we include fermionic matter, compute its energy momentum tensor 
and show how the radiation necessary for a Friedmann cosmology
can be described in our six-dimensional context.
In section 4, we present the effective four-dimensional theory, obtained by
dimensional reduction. There are two massless U(1) gauge fields whose couplings depend 
on $\phi$. In section 5, we present our conclusions.

\section{Cosmological solutions with two symmetric branes}
Six-dimensional Einstein-Maxwell theory couples the metric $g_{AB}$ to a six-dimensional
photon field $A_A$ according to the action
\begin{equation}
 S=\int d^6 x \sqrt{-g}\left \{ -\frac{M_6^4}{2}R + \lambda _6 + \frac{1}{4}F^{AB}F_{AB}
 \right \}.
\end{equation}
Here $g=\det g_{AB}$, $R$ is the six-dimensional curvature scalar, the mass scale $M_6$
can be associated with a six-dimensional Newton constant $G_6=(8 \pi M_6^4)^{-1}$,
$\lambda _6$ is the six-dimensional cosmological constant and the gauge field strength is 
given by $F_{AB}=\partial _A A_B - \partial _B A_A$. The field equations read
\begin{equation}\label{field1}
 R_{AB}-\frac{1}{2}R g_{AB}=M_6^{-4}(T_{AB}^{(F)}+T_{AB}^{(M)}-\lambda _6 g_{AB}),
\end{equation}
\begin{equation}\label{field2}
 \partial _A ( \sqrt{-g}F^{AB})=0.
\end{equation} 
The right hand side of the Einstein equation involves the energy momentum tensor which has
a contribution from the (coherent) photon field
\begin{equation}
 T_{AB}^{(F)}=F_{AC}{F_B}^{C}-\frac{1}{4}F_{CD}F^{CD}g_{AB}
\end{equation}
and from the incoherent fluctuations of radiation or matter $T_{AB}^{(M)}$. Let us first 
investigate the ``vacuum solutions" of the field equations where $T_{AB}^{(M)}$ is
neglected.

We find a class of {\it exact} special solutions with a line element
\begin{equation}\label{metans}
 ds^2 = \exp \left ( -\frac{\phi (t)}{\bar {M}} \right )\{ -dt^2 + a^2(t)
 d \vec{x}d \vec{x}\}+ \exp \left ( \frac{\phi (t)}
 {\bar {M}} \right ) r_0^2 \{ d \rho ^2 + B^2 \sin ^2 \rho \; d \theta ^2 \}
\end{equation} where
\begin{equation}
 r_0^2=\frac {\bar{M}^2}{4 \pi B M_6^4}.
\end{equation}
The gauge field in this solution has only a non-vanishing component in the 
$\theta$-direction \footnote{In a coordinate patch around $\rho = \pi$, one should
use $A_\theta = \frac{m}{2 e_6}(-1-\cos\rho)$.},
\begin{equation}\label{gfsphere}
 A_\theta = \frac{m}{2 e_6} (1-\cos \rho) .
\end{equation}
Here $\vec{x}$ are cartesian comoving three-dimensional coordinates, $\theta$ is a 
periodic angular variable $0 \leq \theta < 2 \pi$ and $\rho$ is in the interval
$0< \rho < \pi$. The field equations (\ref{field1}),(\ref{field2}) 
are obeyed provided
$H(t)=\dot{a}(t)/a(t)$ (dots denote time derivatives) and $\phi (t)$ obey
\begin{equation}\label{cosfi1}
 H^2 = \frac{1}{3 \bar{M}^2}(\frac{1}{2}\dot{\phi} ^2 +V(\phi)),
\end{equation} 
\begin{equation}\label{cosfi2}
 \ddot{\phi}+3H \dot{\phi}+\frac{\partial V}{\partial \phi}=0
\end{equation}
with a potential
\begin{equation}\label{cospot}
 V(\phi)=\bar{M}^4 \left \{ \frac{\lambda _6}{M_6^4 \bar{M}^2}\; e^{-\frac{\phi}{\bar{M}}}
 -4 \pi B \frac{M_6^4}{\bar{M}^4}\; e^{-\frac{2 \phi}{\bar{M}}}+ 2 \pi ^2 m^2
 \frac{M_6^4}{e_6^2 \bar{M}^6}\; e^{-\frac{3 \phi}{\bar{M}}}\right \}.
\end{equation}
We recognize in eqs.(\ref{cosfi1}),(\ref{cosfi2}) the well-known system of 
four-dimensional cosmological field
equations for a homogeneous isotropic universe with a scalar field $\phi$.
For our ansatz (\ref{metans}) we have introduced a Weyl 
scaling factor multiplying the 4D metric in order
to guarantee that the effective 4D Newton's constant $8 \pi G =\bar{M}^{-2}$ is 
independent of time. The appearance of $\phi$ in exponential form is required if $\phi$
is normalized such that its kinetic term takes the standard form. \cite{cw8,sw}

Besides the six-dimensional parameters $M_6$, $\lambda _6$ and $e_6$ our family of
solutions contains several free integration constants: $\bar{M}$ and $B$ as well
as the ``initial values" of $\phi$ and $\dot{\phi}$ at $t=0$ are continuous integration
constants, whereas $m$ is integer. We will see that the integration constant $B$ plays 
a particular role: we will find a family of realistic cosmologies where the effective 
four-dimensional cosmological constant $\lambda _4$ depends continuously on $B$.
Therefore $\lambda _4$ appears as an integration constant of the solution rather 
than being a parameter which is calculable from the six-dimensional couplings. 

Before discussing the properties of $V(\phi)$ and the associated cosmological solutions,
we should understand the geometrical properties of the ansatz.
For fixed $t$ and $\vec{x}$ the internal space spanned by $\rho$, $\theta$ has 
singularities at $\rho =0,\pi$ if $B \neq 1$. In this case it has an ``American football
shaped" geometry with conic singularities at $\rho =0,\pi$ where the deficit angle is
\begin{equation}
  \Delta =2 \pi (1-B).
\end{equation}   
One can interpret this geometry in terms of two branes sitting
at $\rho =0$ and $\rho =\pi$. The brane tension can be directly inferred from the 
bulk geometry.
With a procedure analogous to the one described in \cite{holo},
one finds that the energy momentum tensor $T_{AB}^{(B)}$ of the brane has a 
$\delta$-distribution like form with non-vanishing components only in the
$\mu , \nu$ (4D) direction, 
\begin{equation}
 (T^{(B)})_\mu ^\nu =\frac{B-1}{B r_0^2 e^{\phi /\bar{M}}}\; M_6^4 
 \left (\frac{\delta (\rho)}{\rho}
 +\frac{\delta(\rho - \pi)}{\pi-\rho}\right ) \delta _\mu ^\nu.
\end{equation}
The tension is given by
\begin{equation}
 \mu = - \int _0^\epsilon d \rho \; d \theta \; \sqrt{g_{\rho\rho}g_{\theta\theta}}\; 
 (T^{(B)})^t_t = 2 \pi (1-B) M_6^4.
\end{equation}
This is independent of $\phi$ and corresponds to the energy density of the brane
per four-dimensional volume 
\footnote{Since the brane tension does not appear as an energy density in the
effective 4D theory and is thus a manifestly six-dimensional quantity,
the 4D volume is computed here with the 6D metric $g_{\mu\nu}$, not with the
effective 4D metric $\tilde{g}_{\mu\nu}=g_{\mu\nu}e^{\phi /\bar{M}}$. Otherwise
the tension would indeed depend on $\phi$ and hence on time.}
which is therefore time-independent. However,
we consider the brane tension here as a dynamical quantity, similar in spirit to the
mass of a black hole in the four-dimensional world. In principle, the tension could change 
with time, although it does not for the special solutions considered here. Also similar to
a black hole, the brane never needs to be considered explicitly: the singularities can be 
inferred from the bulk geometry by a type of ``holographic principle" \cite{holo}. Only the
six-dimensional field equations need to be solved.

For the special case $B=1$ the internal space becomes a sphere and no branes are present.
The isometry group is now enhanced to SO(3). For this case the cosmological solutions
of the 6D Einstein-Maxwell theory have been extensively discussed in \cite{cw8}.
Actually, the local solutions with $B \neq 1$ can be mapped to the previously known
local solutions for $B=1$ by the variable change $\theta '=B \theta$, accompanied
by a rescaling $e_6'=Be_6$. (Of course, the periodicity of $\theta '$ differs from $2\pi$
if $B \neq 1$.) By multiplying the gauge coupling in the formulae of \cite{cw8} by a
factor of $B$ they can be taken over directly to the case of a football shaped geometry.

The gauge field configuration (\ref{gfsphere}) 
corresponds to a ``monopole" provided $m$ is an integer.
Indeed, once additional charged fields, e.g. fermions, are introduced, the 
six-dimensional gauge coupling $e_6$ (which has dimension $mass^{-1}$) becomes an
additional free parameter of the action. In addition to the overall mass scale, that we
may take as $M_6$, our model is therefore characterized by two dimensionless quantities
$\lambda _6 /M_6^6$ and $e_6^2 M_6^2$. For the case $B=1$ the monopole configuration
is actually consistent with the spherical symmetry if rotations are accompanied by
suitable gauge transformations \cite{rdss1}.

\subsection{Inflationary Universe}
The character of the possible cosmological solutions depends on the shape of the
potential $V(\phi)$. We observe that $V$ depends on the integration constant 
$B$ which is associated with the deficit angle. In terms of the variable
$x=(M_6^2/\bar{M}^2)\exp(-\phi /\bar{M})$ and the dimensionless parameter combinations
$\tilde{\lambda}=\lambda _6/(4 \pi M_6^6)$ and $\tilde{\mu}=\pi m^2/(2 e_6^2 M_6^2)$,
the potential reads 
\begin{equation}
 V(x)=4 \pi \bar{M}^4 (\tilde{\lambda}x -Bx^2+ \tilde{\mu}x^3).
\end{equation}
For $\tilde{\mu}>0$ and $B^2>3 \tilde{\mu}\tilde{\lambda}$ it exhibits a maximum  at
\begin{equation}
 x_{max}=\frac{B}{3 \tilde{\mu}}\left ( 1-\sqrt{1-\frac{3 \tilde{\mu}\tilde{\lambda}}
 {B^2}}\right )
\end{equation}
and a minimum at
\begin{equation}
 x_{min}=\frac{B}{3 \tilde{\mu}}\left ( 1+\sqrt{1-\frac{3 \tilde{\mu}\tilde{\lambda}}
 {B^2}}\right ).
\end{equation}
The maximum also exists for $\tilde{\mu}=0$ at $x_{max}=\tilde{\lambda}/2B$.
At the maximum $V_{max}=V(x_{max})$
is always positive and we find a solution with an exponential expansion of the scale
factor
\begin{equation}
 a(t)=a_0 \exp (H_{max}t), \quad H_{max}^2=\frac{V_{max}}{3 \bar{M}^2}
\end{equation}
and time independent $\phi = \phi _{max}$. This solution is unstable, and solutions 
starting in the vicinity of $\phi _{max}$ may be associated with the inflationary universe
\cite{cw8}.
However, the curvature of the potential around the maximum is of the order of the
Hubble parameter,
\begin{equation}
 \mu _{max}^2 = \frac{\partial ^2 V}{\partial \phi ^2}|_{\phi _{max}}=-cH_{max}^2,
\end{equation}
\begin{equation}
 c=18 \left ( 1-\frac{1}{2-\frac{B}{\tilde{\lambda}}x_{max}}\right ),
\end{equation}
such that a small deviation $\delta\phi = \phi - \phi _{max}$ grows rapidly on a time
scale set by $H_{max}^{-1}$,
\begin{equation}
 \delta\phi = \delta\phi _0 \exp (\gamma H_{max} t),
\end{equation}
\begin{equation}
 \gamma ^2 +3 \gamma -c=0.
\end{equation}
A sufficient number of e-foldings during inflation would require an extremely tiny 
$\delta\phi _0$. One concludes \cite{cw8} that the present model exhibits the de Sitter
solution suitable for inflation but does not have the quantitative properties for 
inflation to last sufficiently long.

\subsection{Late cosmology with cosmological constant}
The minimum exists only for $m \neq 0$ - the role of the magnetic field is precisely
to lead to a stable ``ground state". Near $\phi _{min}$ the cosmology depends
crucially on the value of the effective four-dimensional cosmological constant
$\lambda _4 =V_{min}=V(x_{min})$.
For $\lambda _4 =0$ we find a static ground state with four-dimensional Poincare 
invariance, i.e. $\phi = \phi _{min}$, $H=0$. For very small $\lambda _4 /\bar{M}^4$
an interesting cosmological solution starts with $\phi (t_0)$ near $\phi _{max}$ in 
an inflationary phase, with $\phi$ then moving towards $\phi _{min}$ and finally 
performing damped oscillations around the minimum. Adding the incoherent fluctuations
of radiation and matter, this corresponds to a quite realistic cosmology !

For arbitrary values of the six-dimensional parameters $\lambda _6$, $M_6$ and $e_6$ 
and arbitrary $m \neq 0$ there exists always a suitable choice of $B=\bar{B}$ such
that $\lambda _4=0$, 
\begin{equation}
 \bar{B}^2 = 4 \tilde{\mu}\tilde{\lambda}=\frac{m^2 \lambda _6}{2 e_6^2 M_6^8}.
\end{equation}
There is therefore always a solution that approaches the 
Friedmann cosmology asymptotically. Indeed, the cosmological solution can be described in
this case by a transition from a de Sitter universe during the inflationary period to
a flat ground state. For $B=\bar{B}$ and with $y=Bx/(2 \tilde{\lambda})$ the 
potential takes the simple form
\begin{equation}
 V=\frac{2 \bar{B}^3 \bar{M}^4 M_6^4 e_6^4}{\pi m^2}\; y(1-y)^2.
\end{equation}
This agrees with \cite{cw8} for $\bar{B}=1$. Whereas the condition $\bar{B}=1$ amounts
to a tuning of the six-dimensional parameters, we have now a whole family of
solutions where $B$ is an integration constant such that the flat space condition
$B=\bar{B}$ can be obeyed for arbitrary values of the six-dimensional parameters. 
Furthermore there are also other solutions with 
$B \neq \bar{B}$ that lead asymptotically to an expansion characterized by a nonzero
cosmological constant $\lambda _4$. For the particular fine tuned choice
$B-\bar{B}\approx 10^{-120}$ one would obtain a realistic cosmology with 
$\lambda _4 =(2 \times 10^{-3}eV)^4$, except for the too short inflation ($c=9/2$,
$\gamma =3(\sqrt{3}-1)/2$).

\subsection{Quintessence Cosmology}
Another type of ``quintessence cosmology" arises if $\phi$ moves away from the 
inflationary solution near $\phi _{max}$ into the direction of large $\phi$, i.e.
$\phi > \phi _{max}$. Without radiation or matter the asymptotic behavior for large
$t$ would correspond in this case to 
\begin{equation}
 H = 2 t^{-1}, \quad \phi = 2 \bar{M}\ln \frac{t}{\sqrt{10} M_6^2 \lambda _6^{-1/2}}.
\end{equation}
In presence of radiation or matter, the same solution is approached.
The fraction of homogeneous ``dark energy" converges to 1,
\begin{equation}
 \Omega _h = \frac{V+ \frac{1}{2}\dot{\phi}^2}{3 \bar{M}^2 H^2}\rightarrow 1,
\end{equation}
already at early times (i.e. not much later than 
a possible entropy production at the end of inflation) and there 
would be no radiation or matter dominated era. At late times the energy density of
the scalar field decreases according to 
\begin{equation}
 V+ \frac{1}{2}\dot{\phi}^2 \propto t^{-2},
\end{equation} 
while matter and radiation decrease as
\begin{equation}
 \rho _m \propto t^{-6}, \quad \rho _r \propto t^{-8},
\end{equation}
respectively. The reason for this unwanted behavior is the precise form of the 
exponential factor in the dominant term of the potential for large $\phi$,
namely exp($-\frac{\phi}{\bar{M}}$). For potentials proportional to exp($-
\frac{\phi}{k \bar{M}}$) with $k< \frac{1}{2}$ exist attractor solutions where the
dark energy density adapts to the incoherent part of the energy density and 
approaches
\begin{equation}
 \Omega _h \rightarrow k^2 n,
\end{equation}
where $n=4$ for radiation and $n=3$ for matter domination. Coefficients $k< \frac{1}{2}$
may be expected to be obtained for the corresponding scalar potential in suitable models with
higher codimensions.

Another problem of these ``quintessence" solutions is that, as we will demonstrate in section 4, 
the present simple model would lead to a substantial time 
dependence of the couplings in the effective four-dimensional theory. This is a direct
consequence of the fact that the action does not exhibit a dilatation symmetry.
Nevertheless we mention the interesting feature that,
independently of all six-dimensional parameters and the values of $m$ or $B$, the
asymptotic value of the dark energy density vanishes. The potential $V(\phi)$ always
approaches zero for $\phi\rightarrow\infty$, and not some positive or negative constant.
In this restricted sense
our model solves both cosmological constant problems, i.e. the asymptotic vanishing
of the energy density for $t \rightarrow\infty$ and the explanation of a tiny
nonzero number $V/\bar{M}^4$. In our model the tiny number is simply a consequence
of the huge age of the universe. 
Finally we note that the ``quintessence" cosmology can also be obtained for $m=0$, i.e. in the
pure six-dimensional Einstein theory without a photon field.

\section{Radiation dominated universe}
Next we want to include radiation and matter into the specific cosmologies described above.
We work in the context of ``holographic branes" \cite{holo} where all particles
correspond to modes with a normalizable tail in the bulk. Then the matter content is
fully specified in terms of the normalizable modes of the six-dimensional bulk geometry.
Again no particular specification of physics on the brane is needed since a type of ``holographic
principle" allows for an extrapolation of the bulk properties to the brane.
We consider the incoherent fluctuations of two 
massless charged six-dimensional Weyl fermions $\Psi$ with opposite abelian charges 
$\pm q$ and equal six-dimensional helicity\footnote{The existence 
of charged particles restricts the allowed monopole numbers such that
$qm$ is an integer.}.
Each of these fermion fields can be harmonically expanded, 
\begin{equation}\label{expansion}
 \Psi (x,\theta ,\rho)=\sum \psi _{ln}(x)\zeta _{ln}(\theta , \rho), 
\end{equation}
where $\zeta _{ln}$ are eigenfunctions of the internal Dirac operator and 
each mode $\zeta$ can be split into a $\theta$ and a $\rho$ dependent part,
\begin{equation}
 \zeta _{ln} = e^{in \theta} \chi _{ln}(\rho).
\end{equation}
After dimensional reduction the eigenvalues of the internal Dirac operator
are seen in four dimensions as mass terms. For most modes the mass is huge of the order
of the compactification scale $M_c \sim r_0^{-1}\exp (-\phi / \bar{M})$. 
At low energies only the massless or very light 
modes are relevant. Unless the existence of light fermions is required by symmetry 
arguments, their appearance would be a pure coincidence and would require some finetuning
of parameters. We therefore assume that the light 4D fermions are precisely the unpaired
chiral modes in the above expansion. These modes are required to be massless in the 
case of unbroken symmetry by virtue of a nonvanishing chirality index \cite{cw6}.
In the football shaped model unpaired chiral modes exist only for fermions that are
charged with respect to the gauge field $A$ \cite{holo}. For $B=1$
they form an irreducible $|qm|$-dimensional representation of the larger 
isometry group SO(3), whereas for $B \neq 1$ the degeneracy may be broken. 
For the energy momentum tensor of these massless modes we expect a relativistic
equation of state. If the U(1)-symmetries are spontaneously broken, 
the fermions typically acquire masses of the order of the symmetry breaking scale.
This would lead to a late time cosmology with
nonrelativistic matter. We neglect spontaneous symmetry breaking here and stick to a 
radiation dominated universe.

Since we are interested in possible
late time cosmologies, we will assume the energy density of these fermions to be small
as compared to the other variables which determine the structure of internal space:
the 6D cosmological constant and the magnetic flux. The fermions can therefore be
considered as a perturbation, with only a tiny backreaction on the shape of internal space.
Furthermore we assume that the characteristic time scale of the cosmological dynamics
is large as compared to $M_c^{-1}$. To a good approximation, 
the internal wave functions of the fermions will then be 
given by the zero modes of the internal Dirac operator on a static football shaped
background with deficit angle $\Delta$. Only the normalization
constant $N_{ln}(t)$ may depend on time 
due to the possible time dependence of the volume of internal space.

In the case of the football shaped model, the zero modes can be computed explicitly. The
internal Dirac equation is
\begin{equation}\label{intdir}
 i \Gamma ^\alpha D_\alpha \zeta(\rho ,\theta)=0.
\end{equation}
Here $D_\alpha$ is the covariant derivative containing the spin connection and the 
gauge coupling, the index $\alpha$ runs over $\theta$ and $\rho$. One gets
\begin{equation}
 D_\theta = \partial _\theta +\frac{1}{2}i \tau _3 (1-B \cos\rho)-ie_6 q A_\theta , \quad
 D_\rho = \partial _\rho  \; .
\end{equation}
The two internal gamma
matrices are defined by $\Gamma ^\alpha = {e^\alpha}_a \tau ^a$, where ${e^\alpha}_a$
is the inverse internal vielbein and $\tau ^1$ and $\tau ^2$ are the first and second Pauli
matrices. This yields
\begin{equation}
 i \Gamma ^\alpha D_\alpha = \left ( \begin{array}{cc}
 0 & -\frac{e^{-i \theta}}{r_0 e^{\phi /2 \bar{M}}}(\partial _\rho -\frac{i}{B \sin\rho}
 D^- _\theta )\\ 
 \frac{e^{i \theta}}{r_0 e^{\phi /2 \bar{M}}}(\partial _\rho +\frac{i}{B \sin\rho}
 D^+ _\theta ) & 0
 \end{array}\right ),
\end{equation} 
where $D^\pm _\theta$ is $D_\theta$ with $\tau _3$ replaced by $\pm 1$.
Consider first the spinor with charge $+q$.
The solutions to eq.(\ref{intdir}) with $\tau _3$ eigenvalue $+1$ are
\begin{equation}\label{iss}
 \zeta ^+ _{0n}(\rho,\theta)=N(t)(\sin \rho)^{-\frac{1}{2}
 +B^{-1}(qm - (n+\frac{1}{2}))}
 (1- \cos\rho)^{B^{-1}(n+\frac{1}{2}-\frac{qm}{2})} 
 \left ( \begin{array}{c}e^{in \theta} \\ 0 \end{array}\right ),
\end{equation} 
where the requirement of normalizability \cite{cw2,holo} constrains $n$ within the limits
\begin{equation}\label{norm1}
 -\frac{B+1}{2}< n < qm + \frac{B-1}{2}.
\end{equation}
The solutions to eq.(\ref{intdir}) with $\tau _3$ eigenvalue $-1$ are similar except
for some sign changes. The normalization conditions require now
\begin{equation}\label{norm2}
 qm - \frac{B-1}{2}< n < \frac{B+1}{2}.
\end{equation}
If $B \leq 1$, only one of the two conditions (\ref{norm1}),(\ref{norm2}) can be fulfilled,
depending on the sign of $m$. 
For the spinor with charge $-q$ the procedure is completely analogous. 
We observe that to each
normalizable mode contained in the spinor with charge $+q$ corresponds a
normalizable mode in the spinor with charge $-q$ with opposite sign of $n$ and 
opposite $\tau _3$ eigenvalue.

We have to compute the six-dimensional energy momentum tensor of
these modes. Since they are massless, we expect a relativistic equation of state,
$p/\rho =1/3$, where $p$ is the pressure in the effective four-dimensional 
world, and $\rho$ is the corresponding effective energy density.
The energy momentum tensor of the fermions is given by the expectation value
\footnote{Usually an energy momentum tensor contains also
a piece $L \delta ^A_B$, where $L$ is the Lagrangian density, but this vanishes here,
since $L=0$ for solutions of the Dirac equation.}
\begin{equation}
 {T^A}_B= \left < \frac{1}{2}i \bar{\Psi}\gamma ^A D_B \Psi \; +h.c. \right >.
\end{equation} 
We assume that the different fermion modes do not mix and have arbitrary phases with
respect to each other, so that all expectation values of the type
$\left < \psi _i^\dagger \psi _j \right >$, $\left < \bar{\psi}_i \psi _j \right >$,
$\left < \psi _i^\dagger \partial _\mu \psi _j \right >$ and
$\left < \bar{\psi} _i \partial _\mu \psi _j \right >$ vanish for different modes,
$i \neq j$. (Here $\psi _i$ corresponds to $\psi _{ln}$ in the expansion (\ref{expansion}).)
Then we can compute the energy momentum tensor for each mode separately.
We therefore take $\Psi (x,\theta ,\rho)=\psi (x)\zeta (\theta , \rho)$, where 
$\zeta = e^{in \theta} \chi (\rho)$
is one particular zero mode.
Furthermore we assume that with each mode also the corresponding antiparticle (with
opposite four-dimensional 
handedness, opposite ``winding number" $n$ and opposite charge $q$ with
respect to gauge field) is excited with the same density, so that no net charges
appear. For each particle contained in one of the six-dimensional Weyl spinors, 
the antiparticle is contained in the other six-dimensional
Weyl spinor.
Every operator $\gamma ^A D_A$ is a tensor product (or a sum of tensor
products) of an operator acting on the 4D part $\psi (x)$ and an operator acting on the
internal part $\zeta (\rho ,\theta)$ of the fermions. We will refer to these as the
4D and the 2D part of the operator. In the absence of warping, a possible choice 
for the 6D gamma matrices is
\begin{equation}
 \gamma ^\mu = e^{\phi / 2 \bar{M}} \; \tilde{\gamma}^\mu \otimes 1, \quad
 \gamma ^{\theta} = {e^\theta}_a \tilde{\Gamma}\otimes\tau ^a, \quad
 \gamma ^{\rho}= {e^\rho}_a \tilde{\Gamma}\otimes\tau ^a .
\end{equation}
Here a tilde denotes the 4D part of an operator and $\tilde{\Gamma}$ is the 4D
$\gamma^5$ matrix. The index $a$ runs from 1 to 2. (In the presence of warping, the 
warp factor would have to be included into the definition of $\gamma^\mu$.) The
covariant derivative $D_\mu$ contains only the 4D part $D_\mu = \tilde{D}_\mu$, i.e. the
2D part is just the identity. (This would not be true in the presence of
warping.) On the other hand, $D_\theta$ and $D_\rho$ have only a 2D part
apart from the factor $\tilde{\Gamma}$.

We assume that the 
distribution of the fermions is homogeneous and isotropic in three-dimensional space.
This forbids any components containing 3D spatial indices except for diagonal 
ones, ${T^{(i)}_{(i)}}$. The effective 4D Dirac equation implies
\begin{equation}
 \bar{\psi}(x)\tilde{\gamma} ^\mu \tilde{D}_\mu \psi (x) =0,
\end{equation} 
where a tilde again denotes a four-dimensional operator. From isotropy then follows
(no summation over $(i)$)
\begin{equation}\label{reos}
 \bar{\psi}(x)\tilde{\gamma} ^{(i)} \tilde{D}_{(i)} \psi (x)=
 -\frac{1}{3}\bar{\psi}(x)\tilde{\gamma} ^t \tilde{D}_t \psi (x).
\end{equation}
In order to connect this to the
6D picture, we compute the 6D energy momentum tensor.
For the diagonal $(\mu\mu)$ components one obtains
\begin{equation}
 {T^{(\mu)}}_{(\mu)}= i \; e^{\phi /2 \bar{M}}\left < \bar{\psi}(x)\tilde{\gamma} ^{(\mu)} 
 \tilde{D}_{(\mu)} \psi (x)\right > |\chi (\rho)|^2.
\end{equation}
Combined with eq.(\ref{reos}) this yields the relativistic equation of state
$p/\rho = {T^{(i)}}_{(i)}/(-{T^t}_t)=1/3$.   
Integrating out the internal dimensions, the effective 4D energy momentum tensor reads
(with an appropriate normalization of $\chi$)
\begin{equation}
 {T^{(\mu)}}_{(\mu),eff}= i \left < \bar{\psi}(x)\tilde{\gamma} ^{(\mu)} 
 \tilde{D}_{(\mu)} \psi (x)\right >.
\end{equation}
This is precisely what one would expect for a massless fermion.

All other components of the 6D energy momentum tensor are zero.
Most terms vanish because $\bar{\psi} (x)\tilde{\Gamma}\psi (x)$ is zero for a 
chiral fermion.
This term occurs in ${T^\theta}_\theta$, ${T^\theta}_\rho$, ${T^\rho}_\theta$ and
${T^\rho}_\rho$, since the 4D part of $\gamma ^\alpha D_\beta$ is just $\tilde{\Gamma}$.
Therefore these components all vanish.
It remains to be shown that the remaining off-diagonal components are also 
zero. We have
\begin{eqnarray}\nonumber
 {T^t}_\rho &=& \left < \frac{1}{2}i \;  
     \bar{\Psi}\gamma ^t D_\rho \Psi \; +h.c. \right > 
    = \left < \frac{1}{2}i \; e^{\phi /2 \bar{M}} \; (\bar{\psi}\tilde{\gamma} ^t \psi ) (
	\chi ^\dagger \partial _\rho \chi ) \; +h.c. \right > \\
  &=& \left < \frac{1}{2}i \; e^{\phi /2 \bar{M}} \; (a^{-1}(t)\psi ^\dagger \psi ) (
	\chi ^\dagger \partial _\rho \chi ) \; +h.c. \right >. 
\end{eqnarray}
It follows from the structure of the internal Dirac equation that
$\chi$ is a real function of $\rho$ times a constant phase (in eq.(\ref{iss}) this phase was
chosen to be 1). Therefore ${T^t}_\rho$  
is purely imaginary and cancelled by the hermitian conjugate. Next
\begin{equation}
 {T^\rho}_t = \left < \frac{1}{2}i \; \bar{\Psi}\gamma ^\rho \tilde{D}_t \Psi
    +h.c. \right >=0
\end{equation}
because of chirality since the 4D part $\bar{\psi}\tilde{\Gamma}\tilde{D}_t \psi$
is zero. A similar argument implies ${T^\theta}_t=0$. 
An interesting role is played by the $t \theta$ component,
\begin{eqnarray}\nonumber
 {T^t}_\theta &=& \left < \frac{1}{2}i \; \bar{\Psi}\gamma ^t D_\theta \Psi \; 
   +h.c. \right > \\ \nonumber
   &=& \left < \frac{1}{2}i \; e^{\phi /2 \bar{M}} \;
    a^{-1}(t)(\psi ^\dagger \psi ) \times \zeta ^\dagger
    (\partial _\theta +\frac{1}{2}i \tau _3 (1-B \cos \rho)-ie_6 qA_\theta )\zeta +h.c. 
    \right > \\
	&=&  - e^{\phi /2 \bar{M}} \; a^{-1}(t) | \chi (\rho)|^2 
	(n \pm\frac{1}{2}(1-B \cos\rho)-e_6 qA_\theta )
	\left < \psi ^\dagger \psi \right >.
\end{eqnarray}
This is real and does not vanish automatically. 
A net charge would indeed lead to a nonvanishing
${T^t}_\theta$. (Such a component would certainly, through Einstein's equations, force one
of the off-diagonal metric components to become nonzero, since ${G^t}_\theta$ 
would identically vanish for a diagonal metric.)  
However we assume here that for each particle the corresponding
antiparticle is also present with equal density. 
This has opposite $n$, opposite $\tau _3$ eigenvalue
and opposite charge $q$ with respect to the 6D gauge field. 
Its ${T^t}_\theta $ component has therefore the same absolute value but the 
opposite sign, and the terms cancel in the total energy momentum tensor. 

The energy momentum tensor can now be inserted into the six-dimensional field
equation (\ref{field1}). Besides the contribution from the massless fermions the (late
time) energy momentum $T^{(M)}_{AB}$ will also contain a similar relativistic
contribution from the two massless four-dimensional gauge fields, the photons.
One observes that the ``internal equations" (for $A,B$ internal indices) are not 
altered. This is reasonable since a change of the shape of internal space corresponds
to the excitation of superheavy modes which should not be affected by a small expectation
value of the fluctuations. The field equation (\ref{cosfi2}) for $\phi$ (corresponding
here to a variable volume of internal space) receives no source term on the r.h.s.
The only modification concerns the equation for the Hubble parameter
$H$ which receives an additional contribution
from the energy density $\rho = -{\tilde{T}^t}_t$. Averaging over internal space
this replaces eq.(\ref{cosfi1}) by the familiar equations
\begin{equation}
 H^2 = \frac{1}{3 \bar{M}^2}(\frac{1}{2}\dot{\phi}^2 + V(\phi)+\rho ), \quad
 \dot{\rho}+4 \rho =0.
\end{equation}   
For the late cosmology with cosmological constant the value of $\phi$ settles at the
minimum of $V$, and therefore $\dot{\phi}\rightarrow 0$, $V(\phi)\rightarrow V_0=\lambda$.
Hence we find a cosmology with radiation and a cosmological constant.

We conclude that 
our scenario has many features required for a realistic cosmology. Starting with a
short inflationary period (if initially $\phi$ is close to $\phi _{max}$), internal
space reaches a stable size and shape (if $\phi < \phi _{max}$), serving as a the ground 
state of the model. The universe becomes radiation dominated and expands with the 
usual Friedmann behavior.

We will see below how the cosmology with the inclusion of the energy momentum tensor
for the massless modes can simply be described by an effective four-dimensional theory.
We close the six-dimensional discussion with the remark that the massive modes
contained in the six-dimensional fields can be included in a straightforward manner.
For these modes the internal Dirac operator leads to a mixing of four-dimensional 
modes with opposite chirality and induces a mass of characteristic size
$r_0^{-1}\exp (-\phi/M)$. For stable particles this mass dominates their contribution
to the energy momentum tensor at late times. However, the massive particles are expected
to decay early in the cosmological evolution and therefore play no role during the
late cosmology of the radiation dominated era.

\section{Dimensional Reduction}
For late cosmology we can concentrate on the massless (or very light) modes.
Then cosmology can be described within an effective four-dimensional framework. Indeed,
in the zero mode ansatz, where all massive modes are neglected, the internal space
may be integrated out. For the above system with relativistic fermions one
obtains the effective four-dimensional Lagrangian
\begin{eqnarray}\label{eff4da}
 L^{(4)}=&-&\frac{\bar{M}^2}{2}R +\frac{Z_1(\phi)}{4}F_{\mu\nu}^{(1)}F^{\mu\nu (1)}
 +\frac{Z_2(\phi)}{4}F_{\mu\nu}^{(2)}F^{\mu\nu (2)}\\ \nonumber
 &+& i \sum _j \bar{\psi}_j \gamma ^\mu (\partial _\mu -i Q_j^{(1)} \bar{e}_1 A_\mu ^{(1)}
 -i Q_j^{(2)}\bar{e}_2 A_\mu ^{(2)})\psi _j 
 +\frac{1}{2}\partial _\mu \phi \partial ^\mu \phi +V(\phi).
\end{eqnarray}
Here the 4D metric $g_{\mu\nu}$ is defined in the Weyl scaled form of eq.(\ref{metans}),
with 6D line element $ds^2=\exp (-\phi /\bar{M})g_{\mu\nu}dx^\mu dx^\nu +...$. 
It determines the 4D Ricci scalar $R=R^{(4)}$ and is used to lower indices.
The scalar potential $V(\phi)$ is given by eq.(\ref{cospot}). The gauge field
$A_\mu ^{(1)}$ is just the dimensionally reduced form of the 6D gauge field $A_\mu$, 
while $A_\mu ^{(2)}$ arises from the internal U(1) symmetry and is proportional 
to the zero mode of the 6D metric components $g_{\mu\theta}$.
The abelian field strength $F_{\mu\nu}^{(i)}$ is defined as usual and the massless 
spinors $\psi _j$ are covariantly coupled to the gauge fields according to their
abelian charges $Q_j^{(i)}$. 
We define the bare four-dimensional gauge couplings $\bar{e}_1$ and
$\bar{e}_2$ such that $Z_1(0)=Z_2(0)=1$. The renormalized
gauge couplings $e_{1,2}$ obey then
\begin{equation}
 e_{1(2)}=\frac{\bar{e}_{1(2)}}{\sqrt{Z_{1(2)}}}.
\end{equation}
We will find 
\begin{equation}\label{gaugerun}
 Z_1=e^{\phi / \bar{M}}, \quad Z_2=e^{2 \phi / \bar{M}}
\end{equation}
and conclude that the renormalized gauge couplings depend on the value of the scalar
field $\phi$. In cosmologies where $\phi$ changes with time, one encounters a time
variation of the ``fundamental constants". In principle, this is a highly interesting
effect which is genuine for quintessence cosmology \cite{cw10,dz}.
In view of the severe restrictions from observation this variation must be very
small, however. Of course, if $\phi$ settles to a fixed value for the ``late
cosmology with a cosmological constant" discussed above, the time variation of the gauge 
couplings becomes unobservable. Except for the possible time variation of the gauge
couplings we find that the effective four-dimensional action (\ref{eff4da}) 
describes a standard four-dimensional theory for massless spinors,
gauge fields and graviton plus a scalar ``cosmon" field.

We want to compute the $\phi$ dependence of $Z_1$ and $Z_2$. Therefore we integrate
the relevant terms in the 6D action. The 6D gauge field
$A_\mu (x,\rho,\theta)$ can be harmonically expanded,
\begin{equation}
 A_\mu (x,\rho,\theta)=\sum \tilde{A}_\mu ^{(lm)}(x)Y^{(lm)}(\rho,\theta),
\end{equation}
where $Y^{(lm)}$ are eigenfunctions of the internal Laplacian. In the special 
case of $B=1$, these are just the spherical harmonics. For the zero mode we write
$\tilde{A}_\mu ^{(1)}(x)$ and omit the indices $(lm)$. 
If needed for the purpose of a clear distinction we mark four-dimensional quantities
by a tilde, e.g. $g_{\mu\nu}^{(6)}=e^{-\phi / \bar{M}}\tilde{g}_{\mu\nu}$. 
As mentioned before, the indices of
four-dimensional fields are raised and lowered with $\tilde{g}^{\mu\nu}$ and
$\tilde{g}_{\mu\nu}$.
The terms in the action containing $\tilde{A}_\mu^{(1)}$ are 
\begin{eqnarray}\nonumber
 S_1 &=& \frac{1}{4}\int d^4x \: d\rho \: d\theta \; (-g)^{1/2}\; F_{\mu\nu}F^{\mu\nu}\\
 \nonumber
 &=&  \frac{1}{4}\int  d^4 x \: d\rho \: d\theta \: (- \tilde{g})^{1/2}\; e^{-\phi / \bar{M}}\;
  r_0^2 B \sin \rho \\ \nonumber
 & \times &\; e^{2 \phi / \bar{M}}\tilde{g}^{\mu\sigma}
  \tilde{g}^{\nu\rho}(\partial _\mu \tilde{A}_\nu ^{(1)}-\partial _\nu \tilde{A}_\mu ^{(1)})
  (\partial _\sigma \tilde{A}_\rho ^{(1)}-\partial _\rho \tilde{A}_\sigma ^{(1)})Y^2 \\
 &=&  \label{act1}\frac{1}{4}\int d^4x (- \tilde{g})^{1/2}e^{\phi / \bar{M}}\; 
   F_{\mu\nu}^{(1)}F^{\mu\nu (1)} \int d \cos\rho \; d \theta \; B r_0^2 Y^2
\end{eqnarray} and
\begin{eqnarray}\nonumber
 S_2 &=& i \int d^4x \: d\rho \: d\theta \; (-g)^{1/2}\:\bar{\Psi}\gamma^\mu (\partial _\mu
   -i q e_6 A_\mu)\Psi \\ \nonumber
 &=& i \int  d^4 x \: d\rho \: d\theta \;(- \tilde{g})^{1/2}\: e^{-\phi / \bar{M}}\;
  r_0^2 B \sin \rho \\ \label{act2}
 & \times & \sum _j \zeta _j^\dagger (\rho,\theta)\zeta _j (\rho,\theta)\;\bar{\psi}_j (x)
  \gamma ^m \tilde{e}^\mu _m \; e^{\phi / 2 \bar{M}}\; (\partial _\mu
  -i q e_6 \tilde{A}_\mu^{(1)}Y(\rho,\theta))\psi _j(x).
\end{eqnarray}
The proper normalization of the kinetic term of the fermions requires
\begin{equation}\label{fermnorm}
 \int d \cos\rho \; d \theta \; e^{-\phi / 2 \bar{M}} B r_0^2 \; \zeta _j^\dagger \zeta _j=1
\end{equation}
for each $j$. Since in $S_2$ the $\phi$ dependence is absorbed by this normalization,
$Y$ has to be independent of $\phi$ in order to lead to a $\phi$-independent $\bar{e}_1$, 
\begin{equation}
 Q_j^{(1)}\bar{e}_1 =e^{-\phi / 2 \bar{M}} B r_0^2 q e_6 \int d \cos\rho \; d \theta \;
 \zeta _j^\dagger \zeta _j Y.
\end{equation}
The normalization $Z_1(0)=1$ yields
\begin{equation}
 \int d \cos\rho \; d \theta \; B r_0^2 Y^2 =1
\end{equation}
and the last line of eq.(\ref{act1}) implies eq.(\ref{gaugerun}). In consequence, 
the renormalized gauge coupling decreases with increasing $\phi$,
\begin{equation}
 e_1 = \bar{e}_1 e^{-\phi / 2 \bar{M}}.
\end{equation}

Next we consider the gauge field corresponding to the invariance of the metric under
translations in $\theta$.
The Killing vector which generates the U(1) symmetry has only a non-vanishing $\theta$
component,
\begin{equation}
 K^\theta = const , \quad K_\theta = const \times \sin ^2 \rho \; e^{\phi / \bar{M}}.
\end{equation}
The gauge field $A_\mu^{(2)}$ is defined by 
\begin{equation}
 g_{\mu\theta}(x,\rho,\theta)=const \times A_\mu^{(2)}(x)K_\theta (t,\rho,\theta)
 =C A_\mu^{(2)}(x) \sin ^2 \rho \; e^{\phi / \bar{M}}
\end{equation}
where the constant $C$ is to be determined. The kinetic term for the second gauge field
is contained in the six-dimensional Ricci scalar,
\begin{equation}
 R_{gauge}=-\frac{1}{4}g^{\mu\rho}g^{\nu\sigma}g^{\theta\theta}(\partial _\mu
 g_{\nu\theta}-\partial _\nu g_{\mu\theta})(\partial _\rho g_{\sigma\theta}
 -\partial _\sigma g_{\rho\theta}).
\end{equation}
Integrating out the internal dimensions for the corresponding part of the 
zero mode action we get
\begin{eqnarray}\nonumber
 S_3 &=& -\frac{M_6^4}{2}\int d^4 x \: d\rho \: d\theta \sqrt{-g}\;R_{gauge}\\
  &=&  \frac{1}{4}\int d^4 x \sqrt{-\tilde{g}}\;F_{\mu\nu}^{(2)}F^{\mu\nu (2)}
  e^{2 \phi /\bar{M}}C^2 \frac{\pi M_6^4}{B}\int _0^\pi d \rho \sin^3 \rho
\end{eqnarray}
and the normalization $Z_2(0)=1$ yields
\begin{equation}
 C^2 = \frac{3 B}{4 \pi M_6^4}.
\end{equation}
The coupling to fermions is given by the term
\begin{equation}
 S_4 = i \int d^4 x \: d\rho \: d\theta \sqrt{-g}\:
 \bar{\Psi} \gamma ^m e_m^\theta D_\theta \Psi .
\end{equation}
Taking $e_m^\theta = -e_m^\mu g^{\theta\theta}g_{\theta\mu}$ and making use of the
normalization condition (\ref{fermnorm}) the integration yields
\begin{equation}
 S_4 = \sum _j\int d^4 x \sqrt{-\tilde{g}}\; \bar{\psi}_j \tilde{\gamma}^\mu A_\mu^{(2)}
 \psi _j \left ( \mp i (n+\frac{1}{2})\right ) \frac{C}{B^2 r_0^2}.
\end{equation}
The charges are
\begin{equation}
 Q_j^{(2)}=\pm (n+\frac{1}{2}),
\end{equation}
where the plus is for left-handed and the minus for right-handed modes. Plugging in the
expressions for $r_0$ and $C$, we find eq.(\ref{gaugerun}) and the effective gauge coupling
\begin{equation}
 e_2=2 \sqrt{3 \pi}\; \frac{M_6^2}{\bar{M}^2}\; e^{-\phi /\bar{M}}.
\end{equation} 
This completes our calculation of the effective 4D action.

\section{Conclusions}
The six-dimensional Einstein-Maxwell theory can lead to an almost realistic cosmology.
We have discussed solutions with a time-varying size of an ``internal space" characterized
by two branes and a ``football-shaped" geometry. With a relativistic energy momentum
tensor induced by the incoherent fluctuations of six-dimensional fermions, gauge fields 
and metric components we have shown how a
Friedmann-Robertson-Walker cosmology arises at late time. The size of internal space plays
the role of a four-dimensional scalar field $\phi$ which is closely associated to
dark energy, either in the form of a cosmological constant
($\phi < \phi _{max}$) or in the form of an exponential potential quintessence
($\phi > \phi _{max}$).

Apart from the many realistic features of our model there remain also considerable problems.
The cosmological constant problem is solved for the quintessence 
solutions with $\phi > \phi _{max}$ in the sense that the ratio of the potential energy
$V$ over $\bar{M}^4$ decreases asymptotically to zero for late time, independently of 
the values of the six-dimensional parameters or the choice of the integration constants
of the solution. However,
in this case the effective 4D gauge couplings are time-dependent beyond the 
observational limits, and the matter density decreases too fast.
The second problem is probably rather easy to solve and the
transition to a realistic radiation dominated cosmology may be achieved in suitably 
modified higher dimensional models. This only requires a different value of $k$
(or a different prefactor for the kinetic term for $\phi$). Stabilizing the running of
the gauge couplings is a challenge - it may be met in higher dimensional theories with
anomalous dilatation symmetry and a fixed point behavior for the running couplings
\cite{cw11}. 

For the solutions with $\phi < \phi _{max}$, the couplings
approach constant values, but here the cosmological constant problem reappears in a 
somewhat different form. For the present solutions an integration constant needs to be
tuned. One may ask if a dynamical adjustment mechanism could favor a small value
of $\lambda _4$, similar to the mechanism of inflation which makes the 3D curvature
small compared to the 4D curvature. 
Furthermore, the period of inflation is in both cases much too short - 
again, this may be different
in other higher dimensional models as already demonstrated in \cite{cw8,sw}.
And finally,
the particle spectrum (i.e. the spectrum of zero modes) is not rich enough
to account for the complicated fermion spectrum that we observe.

The present model may be rendered more realistic by suitable generalizations.
The necessity of tuning an integration constant for the cosmological 
constant for $\phi < \phi _{max}$ may be due to our limitation to only a very special kind
of geometry, namely the ``football shaped" one. A more general ansatz would start with
an arbitrary geometry in the very early universe. It is an interesting speculation
that in the much larger space of general solutions consistent with the symmetries
(i.e. the U(1) isometry, gauge symmetry and invariance under four-dimensional diffeomorphisms) 
a solution with very small 4D cosmological constant is preferred for dynamical reasons, and
is therefore approached for a large class of initial conditions. We will discuss this
issue in a separate paper.

A sufficient inflation may be induced by adding higher curvature terms
\cite{sw}, which are very likely 
to be present as corrections in the effective action since they are generated by
quantum fluctuations.
These higher curvature terms may also be responsible for the compactification of internal space,
instead of the gauge field.
Such a stabilization mechanism becomes important for branes with
codimension higher than two, where the simple monopole configuration does not exist.
    
A more realistic fermion spectrum is obtained when one generalizes the monopole
configuration of the ground state to a non-abelian gauge field. 
A gauge group SO(12) 
instead of U(1) leads to a remarkably realistic fermion spectrum \cite{cw18d1}, with
three generations of quarks and leptons with the appropriate quantum numbers.
In this model one finds even most of the features
required for the observed mass hierarchies and small mixings between the fermions.

Many interesting speculations and discussions invoke unusual features of gravity
which could be generated from higher dimensional physics and modify the late cosmology.
We believe that it is timely to construct realistic higher dimensional cosmologies
that lead to the 
more ``standard" type of a late universe with radiation, matter and dark energy. 
This will establish benchmarks for judging the naturalness of the more drastic modifications.

\end{document}